# Approximation of LRU Caches Miss Rate:

# Application to Power-law Popularities

**Christian BERTHET**
STMicroelectronics, Grenoble, France,

**Abstract**. Building on the 1977 pioneering work of R. Fagin, we give a closed-form expression for the approximated Miss Rate (MR) of LRU Caches assuming a power-law popularity. Asymptotic behavior of this expression is an already known result when power-law parameter is above 1. |It is extended to any value of the parameter. In addition, we bring a new analysis of the conditions (cache relative size, popularity parameter) under which the ratio of LRU MR to Static MR is worst-case.

**Address all correspondence to:** BERTHET Christian; E-mail: Christian.berthet@st.com

## 1. Introduction

Computation of **exact** Miss rate (MR) of a cache with LRU replacement policy has been thoroughly studied. Given an occurrences distribution (so-called 'popularity') of a set of addresses ('alphabet' or 'footprint'), two formulas exist: one proposed by King (1971), also in (Fagin, 1977; Fagin, 1978) and another one by Flajolet et al. (1992). Unfortunately, both of them are intractable and therefore, in practice, one has to resort to **approximation** formulas to evaluate the MR of LRU real-life caches. The following work relies on a very simple approximation stated by Ronald Fagin in 1977.

Let us make a major preliminary point of stressing that, in cache theoretical models and Miss Rate prediction, an underlying hypothesis is always used, namely the 'Independent Reference Model' (IRM). This means that a set of accesses ('trace') respecting the popularity law, is a sequence of independent, identically distributed random variables. IRM is generally not the case in real life cache accesses (in particular for HW Level 1 processor caches) where addresses are subject to some sort of 'clustering' bias; hence, effectively measured Miss Rate is much less pessimistic than the one produced assuming IRM. However, to our knowledge, no one knows how to quantify, in a MR formula, the degree of locality of a trace. So, in the following work, IRM is assumed.

We will focus on popularities that can be modelled as power-laws (a.k.a Generalized Zipf law): items (cacheline addresses for a cache) are ranked according to their popularity (occurrence frequency) and popularity is in a power-law relation with the rank. These laws have since long been recognized as the most accurate way to represent sw-cache interactions (Voldman, 1983) and more generally computer programs (Zhang, 2009).

Let us note also that, in this report, caches are assumed to be fully-associative.
The current report is organized as follows:

Section 2 is a reminder of concepts related to caches and in particular Stack Distance, Working-Set function and Miss Rate.

In Section 3 we introduce Fagin approximation for LRU Miss Rate under IRM and see how it has been recently rediscovered under the "Che's approximation" label.

In Section 4 starting from Fagin equations and moving these equations to the continuous domain, we give an analytic form of LRU MR for power-law popularities under IRM generation, using the Generalized Exponential Integral notation (We use the nickname 'ExpInt').

Section 5 details how this closed-form expression matches previous results from different sources, in particular asymptotics given by Jelenkovic (1999) and Fill (1996). Also we extend Jelenkovic LRU vs Static relation to the case of a power-law parameter between 0 and 1. Static is the '(non-)replacement policy' according to which the D most frequently accessed items are permanently resident in the cache of size D.

Finally, Section 6 develops on the analysis of LRU vs Static MR ratio, and particularly on the maximum of this ratio. Under IRM hypothesis, LRU MR is always worse than the Miss rate of the Static policy. Two quantities are determinant to understand under which conditions this LRU/Static ratio varies: First, the parameter ('exponent') of the power-law, a real positive, and second, the cache ratio $0 \leq \delta \leq 1$ (cache size vs. alphabet size). We found that, for a given parameter 'a' and a given reference alphabet size, there is a cache size (i.e. a cache ratio $0 \leq \delta \leq 1$) for which LRU/static ratio is maximum (i.e. LRU MR is worst-case compared to Static MR). In particular, for a=1 power-law (standard Zipf law), a maximum of 1.43227 is obtained for a cache ratio $\delta=0.453$.

More generally, there is a direct relation between 'a' and $\delta$. In particular, $\delta \rightarrow 1$ when $a \rightarrow 0$ (i.e. when popularity tends to uniform), and $\delta \rightarrow 0$ when $a \rightarrow \infty$.

Although a closed-form expression does not seem to exist to compute the maximum of LRU/static ratio and the corresponding $\delta$, we propose an approximation which appears reasonably good at least for any parameter $a \leq 2$.

We conclude on stressing some open questions. Regarding proofs, the longer ones are given in Appendices. We strived to correlate theoretical results with practical measurements done using different tools: DineroIV (or a variant) for simulation of cache traces, GSL package or WolframAlpha© for mathematical computations.

## 2. Caches: Stack Distance and Re-reference Probability

### *a. Terminology*

We consider traces represented by a sequence of memory accesses to a memory space (also called the alphabet or footprint of the execution). Such a trace results from the execution of programs on a processor and is the input stream to a cache. Each trace is characterized by its length L, and a number N of distinct addresses.

The following terminology to characterize a trace is used:

- An *occurrence* is an access to a specific address. Addresses are characterized by the number of occurrences in the input stream to a cache. Distribution of occurrences among the addresses is often called popularity.
- A *re-reference distance* is the number of accesses (possibly 0) between two accesses to the same address. The pdf (probability density function) of the distribution of re-reference distances is noted $p_{reref}$. The CCDF (complementary cumulative distribution function) of re-reference distances is noted $P_{reref}$.
- A *stack distance* is the number of unique addresses (possibly 0) between two accesses to the same address. The CCDF of the distribution of stack distances is noted $P_{stack}$. Note the re-referenced datum is excluded from stack distance count.

Intuitively, the performance of an LRU Cache depends on the addresses occurrence law of the input stream as well as on the re-reference profile. This observation led to many works on the so-called "Stack Distance" analysis which date back to the 1970s with the seminal paper on Stack Distances (Mattson, 1970).

### b. Working-Set Function

The working set function WS (D) of a trace of accesses is the **average number of distinct** (unique) addresses in a D-size window, D varying from 1 to L, length of the trace.

Obviously WS(1)=1 and the limit of WS(D) when D increases is the number of distinct addresses accessible in the trace under consideration. Also, WS(D)≤D.

When L and N are large, there is no other way to represent WS function than a $\log_{10} \log_{10}$ graph such as the following one. It gives two examples of WS functions for two different power-law popularities (with exponents a=0, i.e., a uniform law, and a=1, i.e., a standard Zipf law) on a 256K state space and traces in the range of 100M generated under the IRM hypothesis.

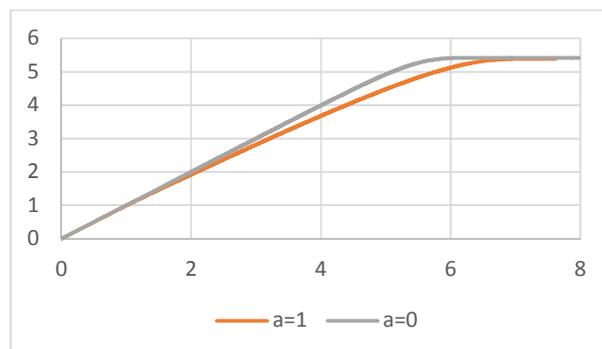

### c. Working-Set Steady-State Relation

Stack Distance analysis consists generally in computing an histogram of the stack distances (whose complexity is in N*log(M) for a sequence of N accesses over M unique addresses).

Rather, the WS(D) function can be computed from the re-reference probability obtained thru a linear traversal of the trace and the following observation:

When D increments by one item, WS(D) increases by the probability that no re-reference of length less than D occurs in the window due to the additional item.
Hence $WS(D+1) = WS(D)*(1-P_{reref}[D]) + (WS(D)+1)*P_{reref}[D] = WS(D)+P_{reref}[D]$.
Remember our definition of CCDF $P_{reref}[D]$ means that D or more accesses sit between two references to the same address. Assuming WS(0)=0, the relation holds for D=0, since WS(1)=WS(0)+$P_{reref}$[0]=0+1=1. It follows that $WS(D) = \sum_{X=0}^{D-1} P_{reref}[X]$.

Another form of this relation is $\Delta WS(D) = P_{reref}[D]$ and is known as Denning and Schwartz's difference equation (Denning, 1972).

### d. Steady-State compared to Trace Sliding window

Following figures are experimental measurements for two traces: I0 L1 (Instruction Level 1 cache) and L2 (Level 2 cache) illustrating the two possible ways to compute the average stack distance.

Traces are respectively 1.8G long for I0 L1 and 83M for L2 and are generated from a real-life platform. For each graph, the blue curve uses the steady-state relation (with $P_{reref}$) while the red one computes directly on the traces the average WS(D) for a sliding window of width D.

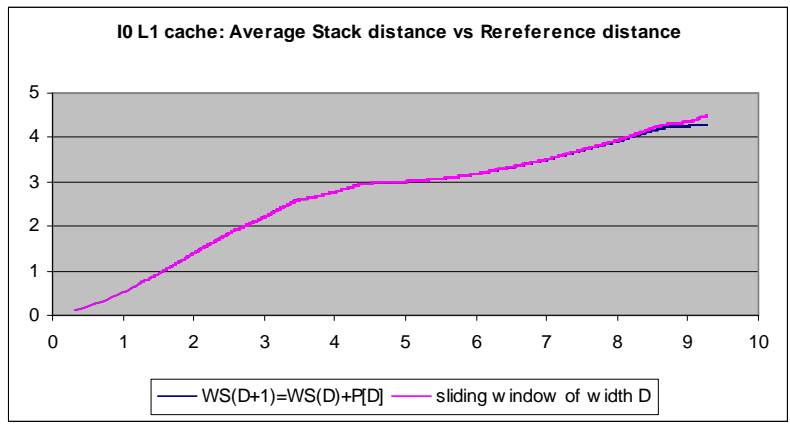

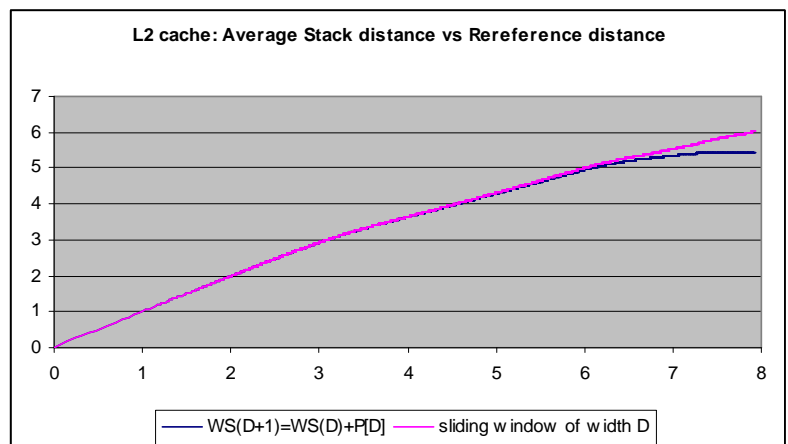

Computation of the red curve is done as follows: The average number of distinct addresses in the incoming stream is evaluated for a moving window of size D using a 0.001 log binning. The algorithm takes ~25min (C code) on a linux 70GB box for L2, but it takes much longer, in the range of 40h, for I0 L1 trace.

Note that both curves match almost entirely, for example on L2 they show two thresholds at approximately $10^3$ and $10^6$. A divergence occurs after $10^8$ for L1 and $10^7$ for L2, so when window width is above 1/10 of total trace length. The difference is due to the fact that the first curve describes the steady-state from the Re-reference probability function, whereas the other one is a measure restricted to the selected trace.

For our purposes, the steady-state relation and corresponding computation of the WS function is perfectly suited to our needs.

### e. Relation Pstack and LRU Miss Rate

$P_{reref}[X>=D]$ is the CCDF probability that at least D records occur between two occurrences of the same cache line address. $P_{stack}[X>=D]$ is the CCDF probability that at least D **unique** records occur in the interval.

LRU replacement policy means that there is a Hit in a cache of size D for an aceess to a given address (i.e., address is in the cache at that time) if and only if there has been accesses to at most (D-1) other addresses since the previous occurrence of the address under consideration. Conversely, an access results in a miss if and only if at least D unique items have been accessed since the previous occurrence of the address.

Hence, for any trace, $P_{stack}$ CCDF calculated at value D is obviously the same expression as the Miss Rate of a LRU cache of size D:    $MR(D)=P_{Stack}[D]$.

## 3. LRU MR Asymptotic under IRM model

### a. $P_{reref}$ and WS function under IRM model

We consider an IRM (Independent Reference Model) framework for each trial of the sequence, in other words every reference is an i.i.d. random variable. Also we assume addresses respect a given popularity distribution $p_i$ among N addresses with $\sum_{i=1}^{N} p_i = 1$.

With these two assumptions, re-reference distance PDF is: $p_{reref}[D] = \sum_{i=1}^{N} p_i (1-p_i)^D p_i$.

In particular, $p_{reref}[0] = \sum_{i=1}^{N} p_i^2$. It is direct that probability is well-formed:

$$\sum_{D=0}^{+\infty} \sum_{i=1}^{N} p_i (1-p_i)^D p_i = \sum_{i=1}^{N} p_i (\sum_{D=0}^{+\infty} (1-p_i)^D) p_i = \sum_{i=1}^{N} p_i \frac{1}{1-(1-p_i)} p_i = 1.$$

Re-reference CCDF (Complementary Cumulative Distribution Function) is:
$$P_{reref}[D] = \sum_{X=D}^{+\infty} \sum_{i=1}^{N} p_i (1-p_i)^X p_i = \sum_{i=1}^{N} p_i (1-p_i)^D (\sum_{X=0}^{+\infty} (1-p_i)^X) p_i = \sum_{i=1}^{N} p_i (1-p_i)^D$$

In particular, $P_{reref}[0]=1$ and $\lim_{D \to +\infty} P_{reref}[D]=0$.

From its definition, the working-set function is:
$$WS(D) = \sum_{X=0}^{D-1} P_{reref}(X) = \sum_{X=0}^{D-1} \sum_{i=1}^{N} p_i (1-p_i)^X = \sum_{i=1}^{N} (1-(1-p_i)^D).$$

WS function verifies WS (0) =0, WS (1) =1 and $\lim_{D\to+\infty}$ WS (D) =N.

### *b. Fagin approximation "in a certain asymptotic sense"*

In (Fagin 1977), a tractable computation of LRU MR is given and is shown to be "correct in a certain asymptotic sense". $P_{reref}$ CCDF is noted M (page 224) and called *expected working-set miss ratio* (parameter is the window size). Another quantity noted S is the *expected working-set size* which is our WS function.

(Fagin 1977) main claim is that "in a certain asymptotic sense", the Miss Rate of an LRU cache of size D (called *expected working-set miss ratio with expected working-set size)* is: $M(S^{-1}(D))$, where $S^{-1}$ is the inverse function of S.
A proof is given as well as measurements for a Zipf's popularity, that justify the claim.

Reminding that $P_{Stack}$ is another notation for LRU MR, Fagin relation is:
$$MR(D)=P_{Stack}[D]=P_{reref}[WS^{-1}(D)].$$
Next figure illustrates Fagin relation for the 83M trace of a L2 cache. It shows that $P_{stack}[X]$ computed by the usual stack distance algorithm is very close to the curve obtained by computing $P_{reref}[WS^{-1}(X)]$. For comparison, we show also $P_{reref}[X]$ as well as the Miss Rate curve obtained using DineroIV tool (Dinero) with LRU-Fully-Associative settings, this curve fits almost exactly $P_{stack}$.

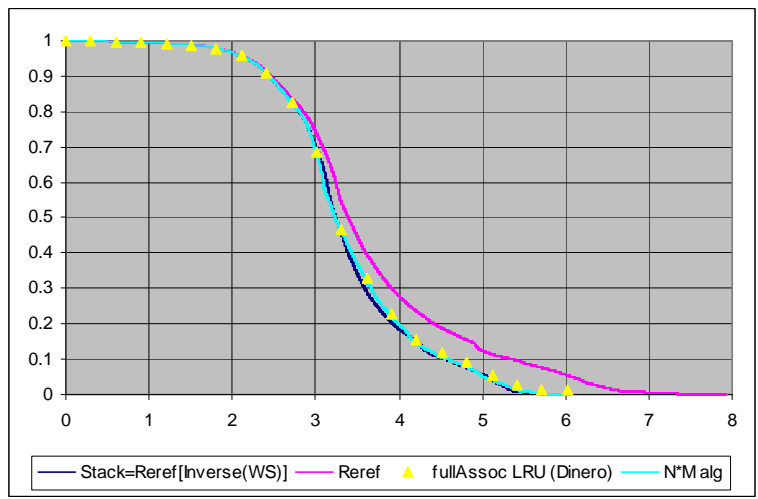

### *c. Constraint on popularity distribution*

We consider the extension to the continuous domain of integer variable D, then $P_{reref}[D]$ and WS(D) are considered as real functions of a positive real variable D.
We also consider a constraint on the occurrences, namely we assume that $p_i \ll 1$ for all i, hence $\ln(1-p_i)=-p_i$. Under this assumption, it is direct that $P_{reref}[D]=WS'(D)$ since:
$$WS'[D] = \sum_{i=1}^{N} -\ln(1-p_i) \cdot (1-p_i)^D = P_{reref}[D]$$

In (Xiang, 2013), this relation is presented as an additional order compared to the usual pdf/CCDF relation: $P_{reref}'[D] = p_{reref}[D]$.

Using Fagin asymptotics equation together with the relation on derivative of inverse function: $\frac{\partial f}{\partial x} \circ f^{-1}(x) = 1 / \frac{\partial f^{-1}}{\partial x}(x)$, one finally obtains:

$$MR[D] = P_{reref}[WS^{-1}(D)] = WS'[WS^{-1}(D)] = 1/[WS^{-1}(D)]'$$

### d. Fagin rediscovered: "Che's approximation"

Another direct consequence of the constraint is that if $\ln(1-p_i) \sim -p_i$ holds for all i, then $\Pr eref[D] = \sum_{i=1}^{N} p_i e^{-p_i D}$ and $WS(D) = \sum_{i=1}^{N}(1 - e^{-p_i D})$.

(Che & al., 2002) introduced an approximation of LRU-FA Hit Rate which is defined as follows (see also (Fricker & al. 2012) ): hit rate of object n of popularity q(n) in a cache of size C is approximated by $h(n)=1-e^{-q(n)\tau}$ where $\tau$ is the unique root of $\sum_{n=1}^{N}(1 - e^{-q(n)t}) = C$.

Using our notation, it is obvious that $\tau$ is the solution of $WS(\tau) = C$ and then, MR of C-size cache is: $\sum_{i=1}^{N} q(i)(1 - h(i)) = P_{reref}[\tau] = P_{reref}[WS^{-1}(C)]$ which is Fagin approximation.

Consequently, "Che's approximation" is essentially a 25 years old re-phrasing of Fagin asymptotic formula together with the constraint $\ln(1-p_i) \sim -p_i$ for each i.

## 4. LRU MR Approximation for Power-law Occurrences

We now consider that popularity is distributed according to a power law, also called Generalized Zipf law in the literature. As usual, let the law be $p_i = k/i^a$, where i, $1 \leq i \leq N$, is the rank of the item, N the size of the addresses footprint ('alphabet'), and k the normalizing constant: $1/k = \sum_{i=1}^{N} \frac{1}{i^a} = H_{N,a}$ ($H_{N,a}$ is the N-th generalized harmonic number).

Preref[D] and WS(D) are calculated by integration on the [0,N] domain of the previous approximations extended to the continuous domain.

### a. Uniform (a=0) Distribution

The pdf form of occurrence probability is $p_i = 1/N$ and its CCDF form: $P_{occur}[D]=1-D/N$. Then, $\Pr eref[D] = \int_{0}^{N} \frac{1}{N} e^{-\frac{D}{N}} dx = e^{-\frac{D}{N}}$ is the exponential distribution with Mean N and $WS(D) = \int_{0}^{N}(1 - e^{-\frac{D}{N}}) \cdot dx = N \cdot (1 - e^{-\frac{D}{N}})$. Note that: $\lim_{N \to \infty} WS(1) = \lim_{N \to \infty} N \cdot (1 - e^{-\frac{1}{N}}) = 1$.

Then for D<N, $WS^{-1}(D) = -N \cdot \ln(1 - \frac{D}{N})$ and finally: $PStack[D] = 1 - \frac{D}{N}$.

In other words, $P_{Stack}$ (i.e. LRU Miss Rate) is also uniform and is the exact replica of $P_{occur}$ as illustrated on next Figure (abscissa is $\log_{10}D$) for $N=2^{10}$, where the Re-reference CCDF $P_{reref}$ is given as well.

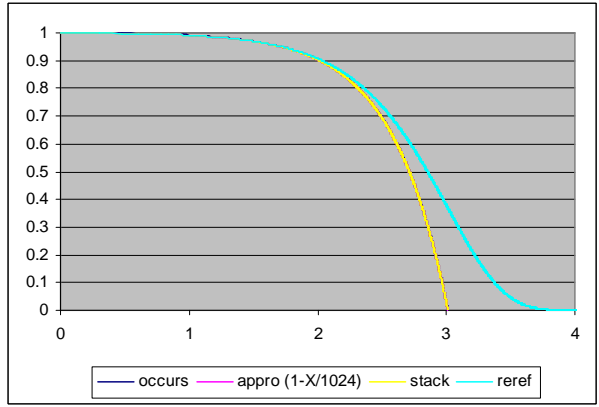

Preref [N] = $e^{-1}$ = 0.367879 and Preref[10*N] = $e^{-10}$ = 4.53997e-05, hence very close to 0.

## b. Zipf (a=1) Distribution

Using the upper incomplete gamma function $\Gamma(a, x)$ (http://dlmf.nist.gov/8.2), following expressions are obtained:

$$\Pr eref[D] = \int_0^N \frac{k}{x} \cdot e^{-\frac{k}{x}D} \cdot dx = k\left[\Gamma(0, \frac{Dk}{x})\right]_0^N = k \cdot \Gamma(0, \frac{Dk}{N}) \text{ since } \Gamma(n, +\infty) = 0, \forall n.$$

and $WS(D) = \int_0^N (1 - e^{-\frac{k}{x}D}) \cdot dx = \left[x - kD \cdot \Gamma(-1, \frac{kD}{x})\right]_0^N = N - kD \cdot \Gamma(-1, \frac{kD}{N}).$

From the relation $\Gamma(s+1, x) = s \cdot \Gamma(s, x) + x^s e^{-x}$, $WS(D) = N + kD \cdot \Gamma(0, \frac{kD}{N}) - Ne^{-\frac{kD}{N}}$ and since $\lim_{x \to +\infty} \Gamma(s, x) \to x^{s-1} e^{-x}$, we see that WS(D) tends to N as expected.

Rather than the incomplete Gamma function we use a more compact form: $E_p$, the generalized exponential integral of order p (http://dlmf.nist.gov/8.19#E1) is related to the upper incomplete gamma function by $E_p(z) = z^{p-1} \cdot \Gamma(1-p, z)$, in particular $E_1(z) = \Gamma(0, z)$
See Appendix 1 for a graph and asymptotic of this function, nicknamed 'ExpInt'.
With this function and introducing $k = 1/H_N$, where $H_{N,1} = H_N = \ln N + \gamma + O(1/N)$, and $\gamma$ is Euler-Mascheroni constant ($\gamma \approx 0.5772$), it stands that:

$$\Pr eref[D] = \frac{1}{H_N} \cdot E_1(\frac{D}{NH_N}) \quad \text{and} \quad WS(D) = N \cdot (1 - E_2(\frac{D}{NH_N})).$$

Note that since $E_p(0) = 1/(p-1)$ for p>1 (http://dlmf.nist.gov/8.19.E6) then WS(0)=0, and it can be shown that: $\lim_{N \to \infty} WS(1) = \lim_{N \to \infty} N \cdot (1 - E_2(\frac{1}{NH_N})) = 1$.

Also $\frac{\partial E_p}{\partial x} = -E_{p-1}$ (http://dlmf.nist.gov/8.19.E13) implies WS'=$P_{reref}$ relation is preserved.

The inverse function of WS is: $WS^{-1}(D) = NH_N \cdot E_2^{-1}(1 - \frac{D}{N})$ for D<=N.

And, finally, we have the closed-form expression: $\boxed{MR[D] = \frac{1}{H_N} E_1(E_2^{-1}(1 - \frac{D}{N}))}$.

Following graph (with $\log_{10}$ abcissa) shows the Occurences (for $N=2^{15}$ addresses), Re-references and Stack CCDFs for a simulation of ($10^3*N$) accesses generated randomly (IRM) according to an occurrence power-law with a=1 parameter.

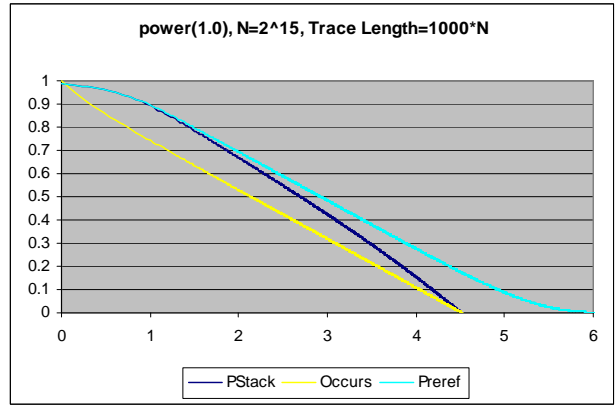

We can now see how previous computations of $P_{reref}$ and $P_{stack}$ based on ExpInt functions fit with these simulations.

Following graphs of $P_{reref}$ and $P_{stack}$ are computed with GSL package (using gsl_sf_gamma_inc incomplete upper gamma function). Strikingly, it appears the approximation is almost perfect for a Distance above 10. Below this value, $\ln(1-p_i) \sim -p_i$ assumption is likely too strong. However, for $P_{stack}$ (and MR) we are interested in much larger cache sizes, hence the assumption is essentially OK for our purposes.

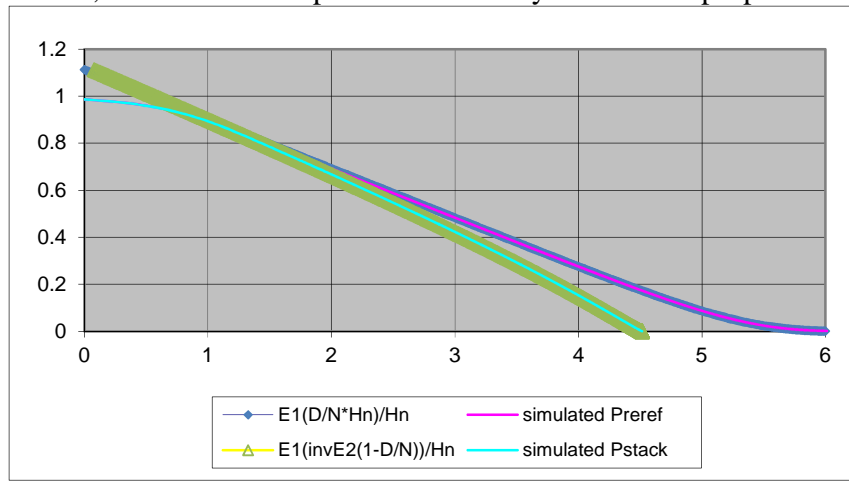

### c. General Power Law (a≠1) Distribution

Preref and WS expressions are:

$$\Pr eref[D] = \int_0^N \frac{1}{H_{N,a} x^a} \cdot e^{-\frac{1}{H_{N,a} x^a} D} \cdot dx = \frac{(D/H_{N,a})^{\frac{1}{a}}}{aD} \left[ \Gamma(\frac{a-1}{a}, \frac{D}{H_{N,a} x^a}) \right]_0^N = \frac{(D/H_{N,a})^{\frac{1}{a}}}{aD} \Gamma(\frac{a-1}{a}, \frac{D}{H_{N,a} N^a})$$

and:

$$WS(D) = \int_0^N (1 - e^{-\frac{1}{H_{N,a} x^a} D}) \cdot dx = \left[ x - \frac{(D/H_{N,a})^{\frac{1}{a}}}{a} \cdot \Gamma(-\frac{1}{a}, \frac{D}{H_{N,a} x^a}) \right]_0^N = N - \frac{(D/H_{N,a})^{\frac{1}{a}}}{a} \cdot \Gamma(-\frac{1}{a}, \frac{D}{H_{N,a} N^a})$$

As in a=1 case, these can be expressed with generalized exponential integral (with a non-integer parameter) $E_p(z) = z^{p-1} \cdot \Gamma(1-p, z)$.

$$\Pr{eref}[D] = \frac{N^{1-a}}{H_{N,a} a} E_{\frac{1}{a}}(\frac{D}{H_{N,a} N^a}) \quad \text{and} \quad WS(D) = N(1 - \frac{1}{a} \cdot E_{1+\frac{1}{a}}(\frac{D}{H_{N,a} N^a})),$$

Consequently, $WS^{-1}(D) = H_{N,a} N^a E_{1+\frac{1}{a}}^{-1}(a(1 - \frac{D}{N}))$. Similarly to a=1, $WS(0) = 0$.

Finally LRU miss rate is: $\boxed{MR[D] = \frac{N^{1-a}}{a H_{N,a}} E_{\frac{1}{a}}(E_{1+\frac{1}{a}}^{-1}(a \cdot (1 - \frac{D}{N})))}$.

Remarks:
There is no discontinuity for a=1 case at this point.
If power-laws parameters are such that a>b, then $WS_a(D) < WS_b(D)$, $\forall D$. Also, when N increases, WS(1) tends to 1.

### *d. Comparison to automatically generated traces*

We use the following code to compute $P_{reref}$ and $P_{stack}$. As in previous section, it uses gsl_sf_gamma_inc GSL function (rather than GSL generalized exponential integral function which unfortunately is limited to integer parameters):

```
double N=pow(2,15); double X; double a=2.0; double inva= 1.0/a; double k=0.0; double Hn=0.0;
int i;
for (i=1;i<=N;i++)  Hn+=1.0/pow(i,a); k= 1.0/Hn; //printf ("%.10f\t%.10f",N, k); printf ("\n");
for (X=0.0; X<=6.0;X=X+0.01) { double D=pow(10,X); double T=k*D/pow(N,a);
    double Y=N*(1.0-(inva*pow(T,inva)*gsl_sf_gamma_inc (-inva, T)));      // Y is WS(D)
    double Z=pow(D*k,inva)*gsl_sf_gamma_inc (1.0-inva, T)/(a*D);          // Z is Preref[D]
    printf ("%.5f\t%.5f\t%.5f\n",X, log(Y)/log(10),Z); }      // Y:WS(X), Z:Preref[X], Y,Z Preref[WS-1()]
```

Then we compare them to graphs of $P_{reref}$ and $P_{stack}$ obtained by simulation (i.e. IRM random generation according to the occurrence power law and computation of $P_{reref}$ and $P_{stack}$ from the trace). Following figures show the results for a=0.5 and a=2. In the former case, match is almost perfect.

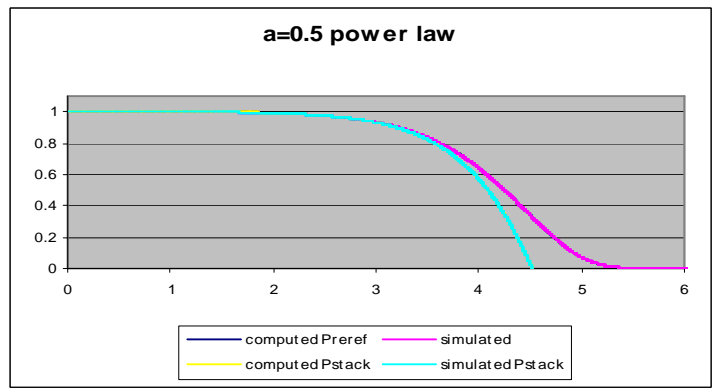

In the latter case, curves are similar for distances above 10. It is not the case below 10 probably for the same reason as before, i.e., logarithm approximation is too strong for high values of exponents.

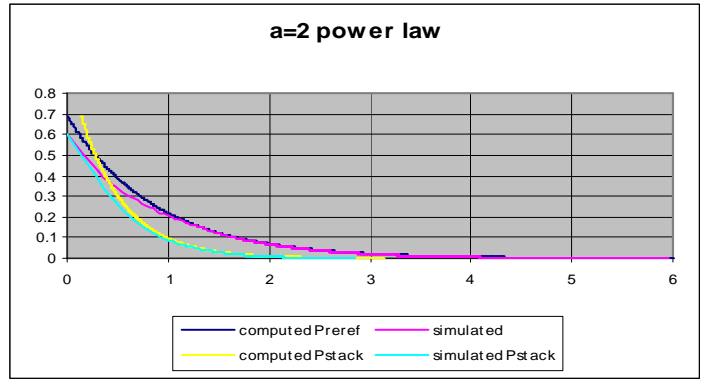

However let us notice that, for values of D corresponding to real-life caches, the match is largely sufficient to guarantee that the approximation is a faithful representation of the $P_{stack}$ CCDF and therefore, the LRU Miss rate.

# 5. Asymptotics of LRU MR Approximation

We first define the Miss Rate for Static caches: In terms of performance, LRU caching scheme is often compared to the static optimal caching.
Then we distinguish between large and small LRU caches since they lead to different expressions regarding the asymptotic behavior when the alphabet size N increases.

### a. Static caches

Static (or 'Static optimal') caching simply consists in 'locking' the most popular D addresses in the cache (in other words, the most frequently accessed items are permanently resident in the cache) (Jelenkovic, 1999). With a minor variant, Static is also called "A0" in (Fagin, 1977). Although it is not necessarily the local optimum, however, optimality is true on a large time window.

Static Miss Rate is the tail of the popularity distribution: $MR_{static}(D) = \int_D^N \frac{dx}{H_{N,a} x^a}$. This leads to the following: If a =1, $MR_{static}(D) = \left[\frac{\ln x}{H_N}\right]_D^N = \frac{\ln N - \ln D}{H_N}$, else

$$MR_{static}(D) = \left[\frac{x^{1-a}}{(1-a)H_{N,a}}\right]_D^N = \frac{N^{1-a} - D^{1-a}}{(1-a)H_{N,a}}$$

This gives (using approximation of generalized harmonic numbers, see Appendix 4) the following asymptotic when N→∞:

If a >1 $MR_{static}(D) \to \frac{1}{(a-1)\zeta(a)D^{a-1}}$, since for a>1, limit of $H_{N,a}$ when N is infinite, is ζ (a), where ζ is Rieman Zeta function.

If a =1 $MR_{static}(D) \to 1 - \frac{\ln D}{\ln N}$, since $H_N$~lnN when N→∞

If 0<a <1 $MR_{static}(D) \to \frac{1}{N^{1-a}}(N^{1-a} - D^{1-a}) = 1 - \delta^{1-a}$, noting δ= D/N.

For a=1/2, Static MR limit is $1-\sqrt{\delta}$. Note the continuity for a=0 with MR=1−δ.

### b. Large LRU Caches

In that case, δ=D/N is close to 1 (case of a large cache size close to alphabet size). Then, parameter a(1-δ) is in the vicinity of 0.
Using ExpInt reciprocal approximation $E_p^{-1}(x) = -\ln(-x\ln x)$ when x→0 (see Appendix 2), and noting it holds regardless of p, it follows that $E_p(E_{p+1}^{-1}(x)) \sim x$.

Hence, for large caches, LRU MR is: $MR(D) \sim \frac{N^{1-a}}{H_{N,a}} a \cdot (1-\delta) = \frac{N^{1-a}(1-\delta)}{H_{N,a}}$

And LRU vs Static ratio is: if a=1: $\frac{(1-\delta)}{H_N}\frac{H_N}{\ln N - \ln D} = \frac{1-\delta}{-\ln \delta}$, otherwise if a≠1:

$\frac{N^{1-a}(1-\delta)}{H_{N,a}}\frac{(1-a)H_{N,a}}{N^{1-a}-D^{1-a}} = \frac{(1-\delta)(1-a)}{1-\delta^{1-a}}$. Therefore LRU to Static ratio does not depend on the exact value of N but rather, on the cache size ratio.

There is a continuity for a=0, i.e. the uniform distribution: in that case both Static and LRU MR are equal to $1-\delta$, hence ratio is 1. Same continuity holds for a=1 since $\lim_{a\to 1}\frac{1-a}{1-\delta^{1-a}} = \frac{1}{-\ln(\delta)}$.

Limit of the ratio is 1 when δ→1: for a=1, obviously from (1-δ)~-ln δ; and for a≠1, from series expansion $\frac{(1-a)(1-\delta)}{1-\delta^{1-a}} = 1 - \frac{a(1-\delta)}{2} + O((1-\delta)^2)$ in the vicinity of 1.

## Asymptotics of LRU MR of large caches when N→∞

When N increases and cache size is large (i.e., δ in the vicinity of 1) LRU MR is:

If 0<a<1, approximation is $H_{N,a} \sim \frac{N^{1-a}}{1-a}$, hence $MR \to (1-\delta)(1-a)$.

Observe this result is coherent with (Fill, 1996) Lemma A.8.a which states that a "density" quantity equivalent to the Hit Rate density (in our representation, it is the opposite of MR derivative) tends to (1-a).

If a>1, $MR \to \frac{N^{1-a}(1-\delta)}{\zeta(a)}$ and, for a =1 (Zipf): $MR \to \frac{(1-\delta)}{H_N} \approx \frac{(1-\delta)}{\ln N}$.

### c. Small LRU Caches

This case (i.e., δ in the vicinity of 0) is much more interesting because it corresponds to real-life caches, with a cache size generally much lower than the alphabet size.

When δ is close to 0, previous approximation on Generalized Exponential integral functions $E_p\left(E_{p+1}^{-1}(x)\right) \sim x$ does not hold.

We first relate our analysis to other models: (Jelenkovic, 1999), (Fricker&al., 2012) and (Fill, 1996).

#### (i) Jelenkovic Asymptotic Relation for a>1

For a>1, (Jelenkovic, 1999) gives an asymptotic formula of LRU cache miss rate compared to 'optimal static' for power-law distributions. LRU Miss Rate Formula when support N increases to infinite, is for a>1: $MR[D] = \frac{1}{a\zeta(a)D^{a-1}}\left[\Gamma(1-\frac{1}{a})\right]^a$, hence LRU to static ratio is $(1-\frac{1}{a})\left[\Gamma(1-\frac{1}{a})\right]^a$. In Appendix 5, we show that this result can be derived simply by using the first two terms of the series expansion of the generalized exponential integral.

An interesting result given by (Jelenkovic, 1999) is that when a→∞, the limit is $e^\gamma$, meaning that LRU MR cannot be worse than ~1.78 times static MR.
Note also that according to Figure 1 of (Jelenkovic, 1999), when power-law parameter approaches 1 (i.e. Zipf law), LRU to static MR ratio tends to 1: LRU is exactly equivalent to static algorithm when support size N→∞.

### (ii) Deriving Jelenkovic 0<a<1 MR formula from Fagin equations

In two different papers (Jelenkovic, 2002) and (Jelenkovic, 2005), Theorem 3 with k=1, the following formula is given: $MR[\partial] = (\frac{1}{a}-1)\eta(\partial)^{\frac{1}{a}-1}\Gamma(1-\frac{1}{a},\eta(\partial))$, where $\eta(\partial)$ is the unique solution of: $1 - a^{-1}\eta^{\frac{1}{a}}\Gamma(-a^{-1},\eta) = \partial$ and $\partial$ the cache size ratio.

Using the generalized exponential integral notation, then $\eta(\partial)$ is simply:

$\eta(\partial) = E^{-1}_{1+\frac{1}{a}}(a(1-\partial))$ and $MR[\partial] = (\frac{1}{a}-1)E_{\frac{1}{a}}(\eta(\partial)) = (\frac{1}{a}-1)E_{\frac{1}{a}}(E^{-1}_{1+\frac{1}{a}}(a(1-\partial)))$. It is very similar to the formula derived from Fagin relation. Indeed they are in exact match to one another assuming approximation of a<1 generalized Harmonic numbers: $H_{N,a} \sim \frac{N^{1-a}}{1-a}$ (see Appendix 4).

### (iii) Fricker&al. formula

(Fricker&al., 2012) Proposition 3 (page 62) gives an asymptotics of a quantity called $t_{\lfloor\delta N\rfloor}$ where $\lfloor\delta N\rfloor$ 0<δ<1 is the cache size, for an un-normalized power law q(n)=1/n^α, 1≤n≤N: $t_{\lfloor\delta N\rfloor} = \psi_\alpha^{-1}(\delta)N^\alpha + o(N^\alpha)$, and $\psi_\alpha$ defined in Lemma 2 for any β>0:

$\psi_\alpha(\beta) = 1 - \int_0^1 e^{-\beta/x^\alpha}dx$. We note that $\psi_\alpha(\beta) = 1 - \left[\frac{\beta^{1/\alpha}}{\alpha}\Gamma\left(-\frac{1}{\alpha},\frac{\beta}{x^\alpha}\right)\right]_0^1 = 1 - \frac{E_{1+1/\alpha}(\beta)}{\alpha}$

Hence $t_{\lfloor\delta N\rfloor} = \psi_\alpha^{-1}(\delta)N^\alpha + o(N^\alpha) = E^{-1}_{1+1/\alpha}(\alpha(1-\delta))\cdot N^\alpha + o(N^\alpha) = \frac{WS^{-1}(\lfloor\delta N\rfloor)}{H_{N,\alpha}} + o(N^\alpha)$

In other words, $t_{\lfloor\delta N\rfloor}$ is asymptotic to WS$^{-1}$ function after normalizing the power law: p(n)=1/( n^α H$_{n,a}$) =q(n)/H$_{n,a}$, 1≤n≤N.

### (iv) Asymptotics for 0<a<1 power law

In (Fill, 1996) detailed results are given for the search cost under the move-to-front rule, problem which has been shown equivalent to the LRU caching (see (Flajolet, 1992)). Formulas are given for the density of the search cost of 0<a<1 power laws (so-called GZL, Generalized Zipf Laws) (see Lemma A.8.b (i), (ii) and (iii)) respectively for a<1/2, a=1/2 and a>1/2) of a quantity noted f$_A$(a) which can be interpreted as the derivative of LRU Hit Rate w.r.t. to the cache size.
We show that these formulas can be derived from the series expansion of the generalized exponential integral function for different cases of 0<a<1.

### 1) Approximation of Reciprocal of $E_{1+1/a}$

$E_{1+1/a}$ expansion in the vicinity of 0 is: $E_{1+\frac{1}{a}}(x) = a + x^{\frac{1}{a}}\Gamma(-\frac{1}{a}) - (\frac{-x}{1-\frac{1}{a}} + \frac{x^2}{2!\left(2-\frac{1}{a}\right)} + ..)$

The trouble is that both Gamma function $\Gamma(-1/a)$ and some denominator of the infinite series may be not defined in the range $0<a<1$, namely for $a=1/2, 1/3, 1/4,....$
If this is not the case, i.e. $a \neq 1/2, 1/3, 1/4,....$, each denominator in the infinite series is defined and since $1/a>1$, the second term is negligible and : $E_{1+\frac{1}{a}}(x) \approx a + x\frac{a}{a-1}$.

In the case $a=1/2, 1/3, 1/4$, etc.. one has to resort to the other series expansion formula (see Appendix 3 on ExpInt singularities). Actually approximation gives the same result:

$E_{1+1/a}(x) = \frac{(-x)^{1/a}}{(1/a)!}(\psi(1+1/a) - \ln x) - \sum_{k=0, k \neq 1/a}^{\infty} \frac{(-x)^k}{k!(k-1/a)} = a + \frac{ax}{a-1} + o(x^2)$

since the first term becomes negligible in the vicinity of 0 when $1/a \geq 2$.

Consequently, a first order approximation of reciprocal function is: $E^{-1}_{1+\frac{1}{a}}(x) = (x-a)\frac{a-1}{a}$

and then $E^{-1}_{1+\frac{1}{a}}(a(1-\delta)) = \delta(1-a)$.

### 2) Approximation of $E_{1/a}$

$E_{1/a}$ series expansion is: $E_{\frac{1}{a}}(x) = \frac{a}{1-a} + x^{\frac{1}{a}-1}\Gamma(1-\frac{1}{a}) - (\frac{-x}{2-\frac{1}{a}} + \frac{x^2}{2!\left(3-\frac{1}{a}\right)} + ..)$

Two cases have to be analyzed depending on whether $1/a-1<1$, i.e. $1>a>1/2$, or not.

### 3) Case $1>a>1/2$

Gamma function is always defined on the interval $]1/2,1[$, thus

$E_{\frac{1}{a}}(x) = \frac{a}{1-a} + x^{\frac{1}{a}-1}\Gamma(1-\frac{1}{a})$. Then $E_{\frac{1}{a}}\left(E^{-1}_{1+\frac{1}{a}}(a(1-\delta))\right) = \frac{a}{1-a} + (\delta(1-a))^{\frac{1}{a}-1}\left(\Gamma(1-\frac{1}{a})\right)$

$\Rightarrow MR[\delta] = 1 - \delta^{\frac{1}{a}-1}(1-a)^{\frac{1}{a}-1}\Gamma(2-\frac{1}{a})$. This matches with (Fill, 1996) lemma.8.b.(iii) result.

### 4) Case $½>a>0$

On the other hand, interval $½>a>0$ includes values for which there are undefined expressions. As described in Appendix ExpInt series expansion singularities (Appendix 3), it can be seen that the factor with the gamma function term can be paired with a term in the infinite series such that the sum is negligible: merging the two terms and passing to the limit leads to an expression which is at least quadratic, so: $E_{\frac{1}{a}}(x) = \frac{a}{1-a} + \frac{ax}{2a-1}$

$$\Rightarrow E_{\frac{1}{a}}\left(E_{1+\frac{1}{a}}^{-1}(a(1-\delta))\right) = \frac{a}{1-a} + \frac{a\delta(1-a)}{2a-1}.$$ And finally $MR[\delta] = 1 - \frac{\delta(1-a)^2}{1-2a}$, matching with (Fill, 1996) lemma.8.b.(i). Note that when power-law parameter 'a' tends to 0 (i.e., a uniform distribution) LRU tends to Static MR, i.e.: (1-δ) as expected.

### 5) Case a=1/2

Point a=1/2 is a singular point. ExpInt series expansion cannot be used directly since $\Gamma(1-\frac{1}{a})$ is not defined. Using approximation $E_2(x) = 1 + x(\ln x + \gamma - 1) + o(x^2)$ and with $E_3^{-1}(\frac{1}{2}(1-\delta)) = \frac{\delta}{2}$, then: $MR[\delta] = E_2\left(\frac{\delta}{2}\right) = 1 + \frac{\delta}{2}(\ln\frac{\delta}{2} + \gamma - 1)$, which is equivalent to (Fill, 1996) lemma.8.b.(ii).

### 6) Case a=1

There again, series expansion cannot be directly used because of undefined terms. See ExpInt series expansion singularities Appendix 3, where are given the series at x=0 of both $E_1(x)$ and $E_2(x)$: $E_1(x) = -(\gamma + \ln x) + x + o(x^2)$ and $E_2(x) = 1 + x(\ln x + \gamma - 1) + o(x^2)$.

Reciprocal function of $E_2(x)$ cannot be easily devised however it can be observed that series expansion of $-E_1(x)/\ln(1- E_2(x))$ tends to $(\ln x+\gamma)/(\ln x+\ln(-\ln x-\gamma+1))$ whose limit is 1 when x tends to 0.

With this limit, $E_1(E_2^{-1}(X)) \approx -\ln(1 - E_2(E_2^{-1}(X))) = -\ln(1-X)$ when X is close to 1. Thus $E_1(E_2^{-1}(1-\frac{D}{N})) \approx -\ln\frac{D}{N}$ and $MR[D] = \frac{1}{H_N} E_1(E_2^{-1}(1-\frac{D}{N})) \approx \frac{\ln N - \ln D}{H_N} \approx 1 - \frac{\ln D}{\ln N}$ when D is close to 0.

MR formula is exactly the static MR for a=1 and is a confirmation of Jelenkovic trend as mentioned on Figure 1 of his paper (Jelenkovic, 1999) which shows a limit of 1 when parameter a tends to 1 (i.e., when N →∞, LRU MR tends to Static MR).

### (v) Relation to other works on Caching analysis

Based on their measurements, (Breslau & al., 1999) claim a ln(D) trend for a=1 power law, and a $D^{(1/a)-1}$ trend for 0<a<1, which are coherent with what we find for 1>a>1/2. Jelenkovic relation for a>1 is also proved (and derived by other means) in (Sugimoto, 2006). (Hattori, 2009) also gives Jelenkovic formula for a>1 laws. They also addressed the case 1>a>1/2 in formula (67) page 18 where they give a result similar to ours where Hit rate is in first order proportional to $t^{1/a-1}$ times a constant proportional to $\Gamma(2-1/a)$: MissRate(t)=1- $\Gamma(2-1/a)$*K* $t^{1/a-1}$ + O(t).

### (vi) Encounter of another kind

A consequence of the WS(D) definition for Power-laws is that the slope at the origin of the WS(D) function in a loglog graph is always 1 when power-law parameter 'a' is 0≤a≤1, and is 1/a when a>1. This observation is proven in Appendix 6.

Not unsurprisingly, this result converges with an empirical observation done in the field of computational linguistics (See formula (7) of (Lü 2010)), and relating the so-called Heaps law (measuring the growth of the vocabulary size width document size) and the power-law frequency distribution of the lexical items (simply named Zipf law in (Lü 2010)).

# 6. Analysis of the Maximum of LRU/Static MR ratio

This Section is intended to be the main contribution of the report. We think the analysis of the maximum of LRU/Static MR ratio is a novelty, bringing confirmation of previous results and opening new questions.

### a. Zipf law (a=1)

LRU/Static MR ratio for Zipf law is for $\delta=D/N$, $1 \leq D < N$: $\dfrac{E_1(E_2^{-1}(1-\delta))}{-\ln\delta}$.

Setting $y = E_2^{-1}(1-\delta)$, ratio is: $F(y) = \dfrac{-E_1(y)}{\ln(1-E_2(y))}$. Following graph of F(y), for $0<y<2$, from WolframAlpha© tool shows a maximum above 0:

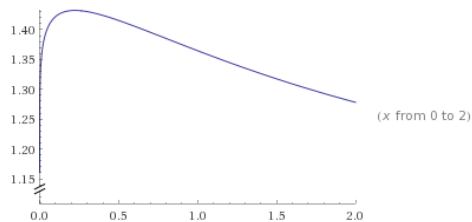

F(y) is close to 1 only on a very small interval above 0 and rises very rapidly (it is 1.14 for y=10⁻⁹ and 1.288 for y=0.001) up to a ~1.43 maximum (for y~0.22) and then tends to 1 as y→∞.

In other words, equality of LRU and Static MR for a Zipf popularity law holds on an extremely small interval. For real caches LRU Miss rate is somewhat higher than Static Miss rate.

It can be shown that F(y) tends to 1⁺ when y →0⁺ (using $E_1$ and $E_2$ series approximations) or y →+∞: using approximation $E_p(x) \sim e^{-x}/x$ which tends to 0 hence $\ln(1-E_2(x)) \sim -E_1(x)$. We can also prove that F(y) is always >1, i.e. LRU MR is always higher than Static MR, in other words, $E_1(y) > -\ln(1-E_2(y))$ for y>0.

This by noting that following relation holds: $f(y) = E_1(y) + \ln(1-E_2(y)) > 0$ since obviously $\lim_{y \to +\infty} f(y) = 0$, $\lim_{y \to 0+} f(y) = +\infty$ using series expansion at origin of $E_1$ and $E_2$ which implies that $f(y) = -(\gamma + \ln(y)) + o(y)$. Its derivative

$f'(y) = -E_0(y) + \dfrac{E_1(y)}{1-E_2(y)}$ is always negative for y>0, since: $E_0(y) = \dfrac{e^{-y}}{y} < \dfrac{1}{y}$ for y>0,

hence $E_0(y)E_2(y) < \dfrac{E_2(y)}{y} = E_0(y) - E_1(y)$, so $E_0(y)E_2(y) - E_0(y) + E_1(y) < 0$ and,

since $0 < E_2(y) < 1$, finally: $-E_0(y) + \dfrac{E_1(y)}{1-E_2(y)} < 0$. Then, F(y) is always above 1.

Function F(y) reaches a maximum when its derivative is null, i.e. when y is solution of $E_1(y)^2 = -(1-E_2(y))\ln(1-E_2(y))E_0(y)$ and $1-\delta = E_2(y)$. Unfortunately, an analytical solution does not seem to exist.

Using WolframAlpha© tool, we found a 1.43227 maximum for y~0.223059 and then a cache ratio δ=1-E₂(0.223059)=0.453.

Coordinates of the maximum (δ=0.453, Max=1.43227) are strikingly confirmed by a set of runs on DineroIV (actually a variant of Dinero allowing for non-power of 2 cache sizes), each run is performed on a 20M IRM trace (over 64K addresses) generated according to a Zipf (a=1) popularity law. Each point is a percent of the cache ratio from 0% to 99%.

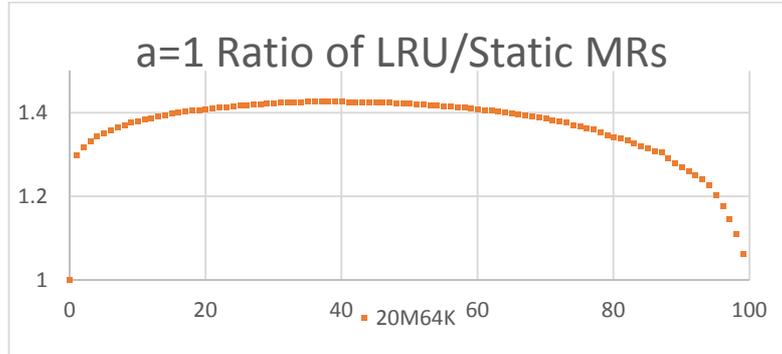

### b. *Generalized power law (a≠1)*

We generalize the ratio LRU/Static MRs to a>0 parameter of the power-law. Ratio for Generalized power law (a>0, a≠1) is (for δ=D/N, 1≤D<N):

$$\dfrac{1-a}{a}\dfrac{1}{1-\delta^{1-a}}\cdot E_{\frac{1}{a}}(E^{-1}_{1+\frac{1}{a}}(a(1-\delta)))$$

Noting p=1/a and $y = E^{-1}_{p+1}(a(1-\delta))$, it is rewritten: $F_a(y) = \dfrac{(p-1)E_p(y)}{1-(1-pE_{p+1}(y))^{1-\frac{1}{p}}}$

### i. Properties of $F_a$ function

Function $F_a(y)$ exhibits the following interesting properties:

**Property P1:** $\lim_{a\to 0+} F_a(y) = 1, \forall y > 0$.

This stems directly from $E_n(x) \sim E_{n+1}(x)$ when n→+∞, and is coherent with a constant ratio equal to 1 for a uniform popularity.

**Property P2:** $\lim_{a\to 1} F_a(y) = F_1(y)$, where $F_1(y)$ is defined in the previous paragraph for a=1: $F_1(y) = \dfrac{-E_1(y)}{\ln(1 - E_2(y))}$. This comes from: $\lim_{p\to 1} \dfrac{(p-1)}{1-(1-pu(x))^{1-\frac{1}{p}}} = \dfrac{-1}{\ln(1-u(x))}$.

**Property P3:** $\lim_{a\to\infty} F_a(y) = \dfrac{E_0(y)}{e^{E_1(y)} - 1}$. We note this function $F_\infty$. This limit comes from: $\lim_{p\to 0} \dfrac{(p-1)}{1-(1-pu(x))^{1-\frac{1}{p}}} = \dfrac{1}{e^{u(x)} - 1}$. It holds that $\lim_{y\to\infty} F_\infty(y) = 1$, and $\lim_{y\to 0} F_\infty(y) = e^\gamma$. One can verify that $F_\infty(y) > 1, \forall y > 0$, from the consideration that $\ln(E_0(y)+1) - E_1(y)$ is always positive, since it tends to 0 when y→+∞, to γ when y→0 and derivative $\dfrac{-E_{-1}(y)}{E_0(y)+1} + E_0(y) = \dfrac{E_0(y)}{E_0(y)+1}\left(E_0(y) - \dfrac{1}{y}\right)$ is always negative.

**Property P4:** $\lim_{y\to +\infty} F_a(y) = 1^+$.

When y→+∞, $\delta = 1 - pE_{p+1}(y) \to 1$, meaning the cache is large. When y→+∞, $z = E_n(y) \to 0$, regardless of n, hence: $\lim_{y\to +\infty} F_a(y) = \lim_{z\to 0} \dfrac{(p-1)z}{1-(1-pz)^{1-\frac{1}{p}}}$. Using Laurent series when x→0: $\dfrac{1}{1-(1-x)^n} = \dfrac{1}{nx} + \dfrac{n-1}{2n} + o(x)$, finally,

$\lim_{y\to +\infty} F_a(y) = \lim_{z\to 0} (p-1)z\left(\dfrac{1}{z(p-1)} + \dfrac{-1}{2(p-1)} + o(z)\right) = 1$.

**Property P5:** $\lim_{y\to 0+} F_a(y) = 1$ when 0≤a≤1, and $\dfrac{a-1}{a}\left(\Gamma(1-\dfrac{1}{a})\right)^a$ when a>1.

Proof is given in Appendix 7. Result for a>1 is the relation given by (Jelenkovic, 1999). In the sequel we note $F_a(0)$ the Jelenkovic limit. It is always above 1, tends to 1 when a→1, and tends to $e^\gamma$ ~1.781 when a→∞. Note it is valid only for small caches (y→0 means cache ratio δ→0).

**Property P6:** $\forall a>0, \forall y>0,\ F_a(y)>1$  Proof is in Appendix 8.

**Property P7:** $F_\infty(y) > F_1(y),\ \forall y > 0$  Proof is in Appendix 9.

**Summary of the Properties**

$F_a(y)$ is a monotonically-increasing family of functions such that, for any y>0:

if 0<a<1<b<+∞ : $F_0(y) < F_a(y) < F_1(y) < F_b(y) < F_\infty(y)$.

### c. Calculation of the Maximum

Derivative of $F_a$, a>0 and a≠1, is null when y is solution of:

$$F_a(y) = \frac{\frac{d}{dy}\left((p-1)E_p(y)\right)}{\frac{d}{dy}\left(1-(1-pE_{p+1}(y))^{1-\frac{1}{p}}\right)} = \frac{-(p-1)E_{p-1}(y)}{\left(-E_p p\left(1-\frac{1}{p}\right)\right)(1-pE_{p+1}(y))^{-\frac{1}{p}}} = \frac{E_{p-1}(y)(1-pE_{p+1}(y))^{\frac{1}{p}}}{E_p(y)}$$

and Maximum of ratio is then $F_a(y_0)$ if such a solution $y_0$ exists. Yet this does not lead to an analytic form of its zeroes.

For a=1 this leads to equation $F_1(y) = \frac{E_0(y)(1-E_2(y))}{E_1(y)}$ which, as we have seen, has a unique solution (see previous section) but does not lead to a closed-form expression.

For a→0, since $\lim_{a \to 0} F_a(y) = \lim_{p \to \infty} \frac{E_{p-1}(y)(1-pE_{p+1}(y))^{\frac{1}{p}}}{E_p(y)} = 1$, this means that for any y, $F_0(y)$ is a maximum equal to 1.

When p→0 (i.e., a→∞), $\lim_{p \to 0} \frac{E_{p-1}(y)(1-pE_{p+1}(y))^{\frac{1}{p}}}{E_p(y)} = \frac{E_{-1}(y)}{E_0(y)e^{E_1(y)}} = \frac{y+1}{ye^{E_1(y)}}$ hence a

maximum is found for y such that $\frac{E_0(y)}{e^{E_1(y)}-1} = \frac{y+1}{ye^{E_1(y)}}$. Each of those two expressions tends to $e^\gamma$ when y→0, therefore the Ratio Maximum sits at abciss y=0 and is equal to $F_\infty(0) = e^\gamma$.

### d. Graph of Maximum

Approximations are made for specific values of parameter 'a' using WolframAlpha© tool.

For example, for a=1/2, ratio $F_{1/2}$ is $\frac{E_2(y)}{1-(1-2E_3(y))^{\frac{1}{2}}}$ with limit $F_{1/2}(0)=1$ and a maximum

at 1.28732 for y=0.55779224, i.e. δ=0.5927514.

For a=1/3, maximum is 1.21657 for y=0.8158 and δ=0.6729.

Similarly a maximum above $F_a(0)=1$ exists for any 0<a≤1. Intuitively, δ increases to 1 as a gets closer to 0.

For a=2, ratio is $\frac{E_{1/2}(y)(2-E_{3/2}(y))}{2E_{3/2}(y)}$ giving a maximum of 1.58 with Jelenkovic limit

$F_2(0) = \frac{1}{2}\left(\Gamma(\frac{1}{2})\right)^2 = \pi/2 \sim 1.57$ very close to the maximum value, while abciss of the maximum is very close to 0 (<0.035).

Extending to a number of values of parameter 'a' and using both WolframAlpha© or GSL, we obtain the following graph which compares the Max of LRU/Static ratio with Jelenkovic limit: as 'a' increases, $F_a(0)$ limit gets closer to Maximum of ratio, and both tend to $e^\gamma$. When a is above 2, the maximum is very close to $F_a(0)$ Jelenkovic limit.

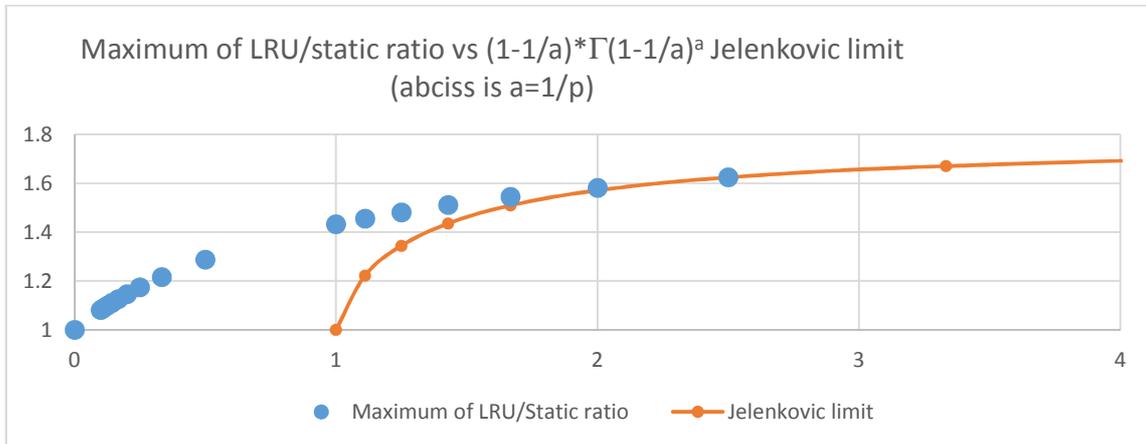

The corresponding cache ratio as a function of power-law exponent 'a' is as follows: $\delta \to 1$ when $a \to 0$ (i.e. popularity is uniform), and $\delta \to 0$ when $a \to \infty$.

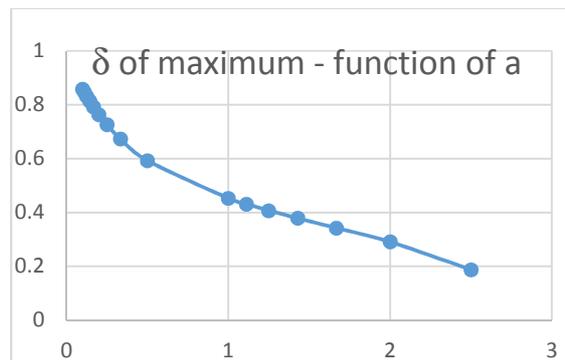

In (Jelenkovic, 1999) Fig. 2 experiments with a=1.4 show that simulation results for a cache ratio below $10^{-3}$ (cache size up to $10^3$ on a vocabulary of size $10^6$) are very close to $F_a(0)=1.42362$ constant approximation of LRU to Static ratio. The figure above comparing Max $F_p$ to $F_p(0)$ shows a ~1.5 maximum when a=1.4, which is reached for $\delta$~0.38, so a cache size much higher than those analyzed in (Jelenkovic, 1999) Fig. 2. This explains why $F_p(0)$ is an excellent approximation of $F_p(\delta)$ for low values of $\delta$.

On the other hand, when parameter a is just above 1, and very close to 1, this approximation may be at risk depending on the size of the cache: we have seen that LRU/Static ratio for a=1 goes from 1 (for $\delta=0$) with a very steep curve up to a maximum of 1.43227 (for $\delta=0.453$). It is so steep that its value is 1.243 for $\delta=0.00096$ (abciss=0.0001). So, for a cache in the range studied by (Jelenkovic, 1999), when parameter a is close to 1, $F_p(0)$ approximation can lead to an underestimation of LRU MR in the range of 25%.

Clearly this underestimation gets smaller as parameter a increases

### e. Confirmation with Cache simulation tools

Following graph shows 99 runs (cache ratio varying from 1 to 99%) on a Dinero-variant tool for a 20M IRM trace generated over 64K addresses according to a power-law (for both a=0.1 and a=0.2). They confirm the results obtained both for the cache ratio of the maximum and the maximum itself.

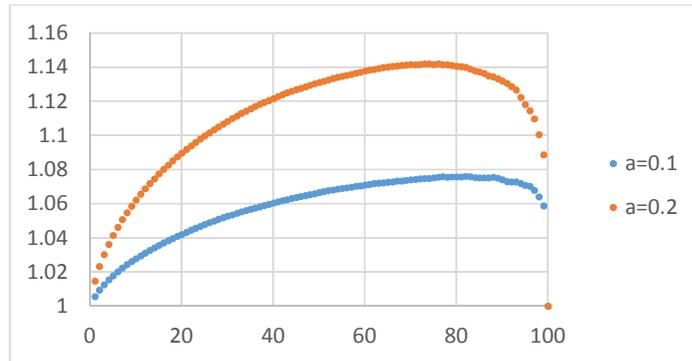

For a>1, we have results on DineroIV tool (hence restricted to power-of-2 caches from 0 to 1K) on 100M traces IRM-generated over 1K addresses for a set of values from a=1.0 to a=2.0. It shows the trend to $F_a(0)$ as well as the very steep slope at the origin. Having in mind that a real cache is in the range of 0.1 to 1% of the address space, this justifies the concern made in the previous paragraph.

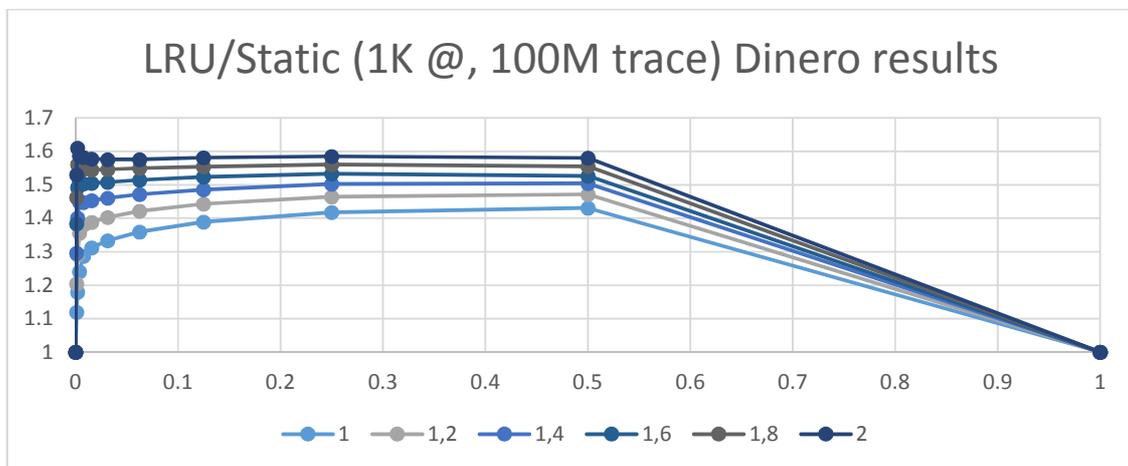

### f. A possible approximation of abciss of Maximum

Repeating previous computations using WolframAlpha© on a number of points, we found the following graph comparing the abciss of the maximum of $F_p$ with the function $E_1(1/p)$ and they appear to match almost exactly up to a=2 (i.e. p≥0.5)

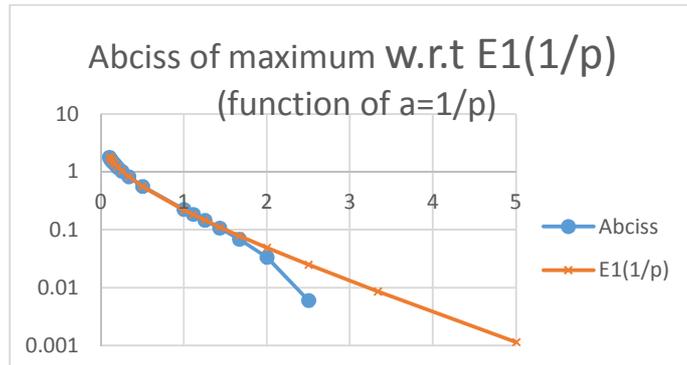

A similar (slightly more precise) graph can be obtained using GSL package. Divergence from $E_1(1/p)$ can be further analyzed above p=2.

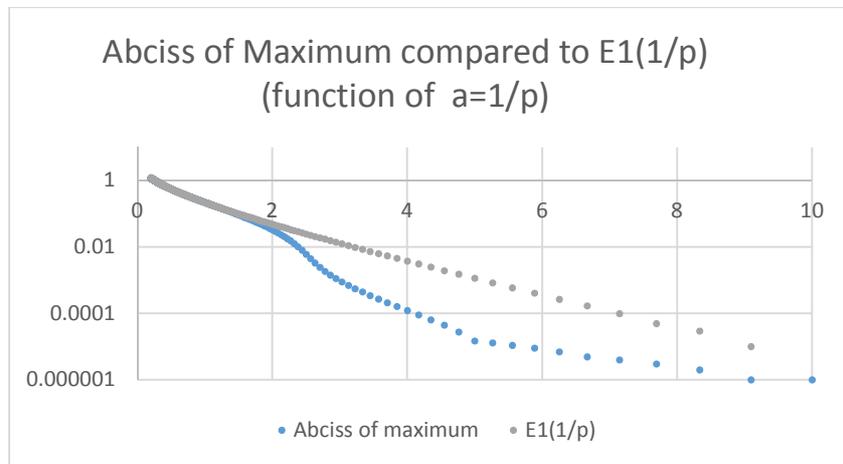

For a=1, using $x=E_1(1)=0.293839$, the value of maximum is 1.43129 which is coherent up to the 3th decimal with the solution of Maximum produced by Wolfram.

In conclusion, $E_1(1/p)$ seems to be an excellent predictor of the abciss of maximum when p>0.5 (i.e. 2>a≥0). Below (i.e a≥2) it is not the case, however we have seen that for these values, maximum is extremely close to the value at origin (Jelenkovic's value).

Finally we conjecture that,

for p>0.5, $y=E_1(1/p)$ is a good approximate solution of the equation:

$$(p-1)(E_p(y))^2 = E_{p-1}(y)\left[(1-pE_{p+1}(y))^{\frac{1}{p}} - (1-pE_{p+1}(y))\right]$$

In particular, using the limits where p=1, $y=E_1(1)=\Gamma(0,1)$ is a good approximate solution of $E_1(y)^2 = -(1-E_2(y))\cdot\ln(1-E_2(y))\cdot E_0(y)$.

## 7. Conclusion

In this report, we have proved that a closed-form expression for power-law popularities can be derived from R. Fagin LRU Miss rate approximation. Asymptotics of this expression are coherent with previously known results.

The main contribution of this work is in a more thorough analysis of the LRU/static ratio which shows that, for any real positive power-law parameter 'a', there is a cache ratio $0 \leq \delta \leq 1$ (cache size vs. alphabet size) for which LRU/static MR ratio is maximum.

This analysis relies on the Generalized exponential integral functions, for which some new properties have been stated and proved.

Solution of the LRU/Static ratio maximum and corresponding cache ratio have been found in some cases, and an approximation is provided when the popularity parameter is less than 2.

However, a closed-form expression of these quantities is still an open problem.

## *Appendix 1: Generalized Exponential Integral: 'ExpInt'*

Graph of Generalized Exponential Integral $E_p(x)$ is given in http://dlmf.nist.gov/8.19#F1

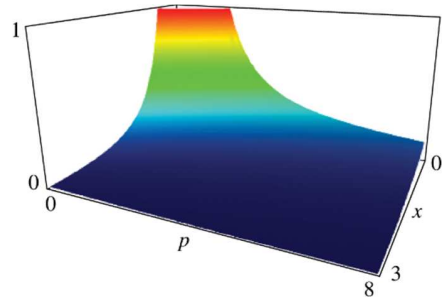

**General considerations**
Maplesoft uses the notation: $Ei(a,z)=z^{a-1}\,\Gamma(1-a,z)$ and WolframAlpha© uses expint(a,z).
We use the nickname 'ExpInt' for Generalized Exponential Integral.

$E_0(x)=x^{-1}e^{-x}$ hence $E_0(x)=1$ for $0<x<1$, solution of $xe^x=1$, which is Lambert W function (aka product log) with $W(1)=0.567143$.

We use relations $E_p(0)=1/(p-1)$, $p>1$ http://dlmf.nist.gov/8.19.E6 and $\dfrac{\partial E_p}{\partial x} = -E_{p-1}$

(http://dlmf.nist.gov/8.19.E13). And also relation $nE_{n+1}(x)=e^{-x}-xE_n(x)$.
We are interested in $E_{1+1/a}$ functions $a>0$. We have $E_{1+1/a}(0)=a$, for $a>0$.
Since $E_2(0)=1$, $p=2$ is a particular point for $E_p(0)$.

**Vicinity of $+\infty$**
Generalized exponential-integral $E_p(x) \sim e^{-x}/x$ when $x\to\infty$, regardless of p. See asymptotic series expansion in
http://functions.wolfram.com/GammaBetaErf/ExpIntegralEi/introductions/ExpIntegrals/ShowAll.html

**Vicinity of 0**
Obviously an approximation of $E_{1+1/a}$ will depend how $1+1/a$ compares to 2, i.e. a to 1.

When $0<a<1$, $E_2>E_{1+1/a}>E_\infty$. Slope of $E_{1+1/a}(x)$ at point $x=0$ is $(-E_{1/a}(0)=a/(a-1)$, hence a possible linear approximation of $E_{1+1/a}$ around 0 is $E_{1+\frac{1}{a}}(x) \sim a - \dfrac{ax}{1-a}$.

When $1\leq a<+\infty$, $E_1>E_{1+1/a}\geq E_2$ and $E_{1+1/a}(0)=a$, therefore $+\infty > E_{1+1/a}(0)>1$.
Slope of $E_{1+1/a}(x)$ at point $x=0$ is $(-E_{1/a}(0))$ which is infinite for $a=1$ since $E_1(0)=+\infty$.
When a increases, $E_{1+1/a}$ function tends to $E_1(z)=\Gamma(0,z)$ function, but starting from an ever-increasing origin since $E_{1+1/a}(0)=a$.

For Generalized Exponential Integral, an interesting reference is (Chiccoli, 1990).

## *Appendix 2: Approximation of Reciprocal of Generalized Exponential Integral*

We apply previous results to an approximation of reciprocal $E^{-1}_{1+\frac{1}{a}}(x)$ when $x \in [0,a]$.

**Vicinity of 0**

It is known that Lambert function W is the inverse of $xe^x$ function.
Hence the inverse function: Inverse($e^{-x}/x$) = W(1/x). This can be checked with WolframAlpha© tool:
http://www.wolframalpha.com/input/?i=inverse+of+1%2F%28xe^x%29
Since $E_p(x) \sim e^{-x}/x$ when $x \to \infty$, regardless of p, this gives the following: $E_p^{-1}(x) \sim W(1/x)$.
It is known that W(x) has a first-order approximation ln(x)-ln(ln(x)):
http://dlmf.nist.gov/4.13
Hence we can approximate the reciprocal of the exponential integral function in the vicinity of 0+, regardless of parameter p, by:

$$E_p^{-1}(x) = \ln(1/x) - \ln\ln(1/x) \text{ or, simply,}$$

$$E_p^{-1}(x) = \ln(1/x)$$

**Vicinity of a**

Obviously vicinity of 0+ cannot extend higher than x=1 (since ln(ln(x)) is not defined after this point) however, approximation may be acceptable over whole [0,a] if a is sufficiently small. If this is not the case, we use an approximation of $E_p(x)$ when x=0.

When 0<a<1, we have the following : $E^{-1}_{1+\frac{1}{a}}(x) \sim \dfrac{a-1}{a}(x-a)$

## *Appendix 3: ExpInt series expansion singularities*

We are interested in the series expansion of ExpInt function in the vicinity of 0:

$$E_p(x) = \frac{1}{p-1} + x^{p-1}\Gamma(1-p) - \left(\frac{-x}{2-p} + \frac{x^2}{2!(3-p)} + ..\right).$$

This series expansion is not defined for a positive integer p since Gamma function Γ(1-p) is not defined at p=1,2,3,.. and also for each of these values of p, there exists a value of k for which denominator is null (respectively for k=0,1,2,..).

However there is an interesting limit that can be computed for each case of p. For p=1:

$$\lim_{p\to 1}\left(\frac{1}{p-1} + x^{p-1}\Gamma(1-p)\right) = \lim_{p\to 0}\left(\frac{\Gamma(p)}{x^p} - \frac{1}{p}\right)$$ where we find the very nice relation

$$\lim_{p\to 0}\left(\frac{\Gamma(p)}{x^p} - \frac{1}{p}\right) = -(\ln x + \gamma)$$ with γ the Euler-Mascheroni constant, hence the

approximation $E_1(x) = -(\ln x + \gamma) + x + o(x^2)$. The nice relation stems directly from the definition of γ : $\gamma = \lim_{p\to 0}\left(\frac{1}{p} - \Gamma(p)\right)$ and the series expansion of Gamma function.

More generally, a known relation can be used for positive integer parameters of ExpInt function http://dlmf.nist.gov/8.19.E8 :

$$E_n(z) = \frac{(-z)^{n-1}}{(n-1)!}(\psi(n) - \ln z) - \sum_{\substack{k=0 \\ k\neq n-1}}^{\infty} \frac{(-z)^k}{k!(1-n+k)},$$

Which gives $E_1(x) = (\psi(1) - \ln x) + x + o(x^2)$ where ψ is the digamma function such that ψ(1)= - γ, so $E_1(x) = -(\gamma + \ln x) + x + o(x^2)$

Using ψ(n)- ψ(n-1)=1/(n-1), we have ψ(2)= - γ+1 and
$$E_2(x) = -x(\psi(2) - \ln x) + 1 + o(x^2) = 1 + x(\ln x + \gamma - 1) + o(x^2).$$

Note that correctly $E_2'(x) = -E_1(x)$. Similarly
$$E_3(x) = \frac{x^2}{2!}(\psi(3) - \ln x) + \frac{1}{2} - x + o(x^3) = \frac{1}{2} - x + \frac{x^2}{4}(-2\gamma + 3 - 2\ln x) + o(x^3).$$

For an ExpInt parameter n above of equal to 3, singularity (ψ(n)-lnx) is applied to a monomial of exponent at least 2. This form is given by WolframAlpha© under the name Generalized Puiseux series (see http://functions.wolfram.com/GeneralIdentities/4/ ).

## *Appendix 4: Generalized harmonic numbers*

The Nth generalized Harmonic number is $H_{N,a} = \sum_{i=1}^{N} \frac{1}{i^a}$.

For N=28 and 0<=a<=2, the following graph is obtained:

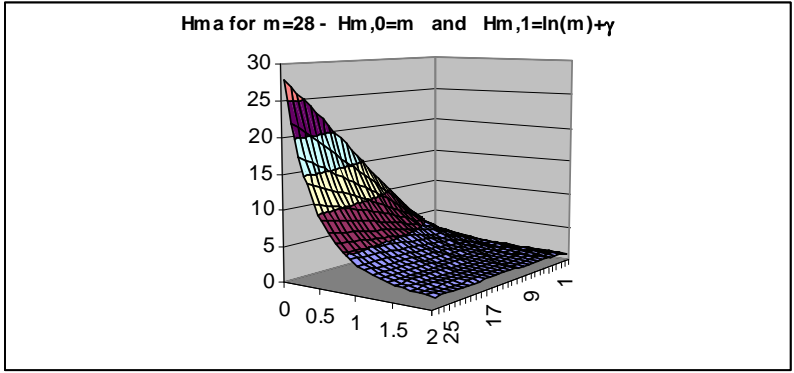

If a =1, it is well known that $H_N = \ln N + \gamma + O(1/N)$ where $\gamma$ is the Euler-Mascheroni constant ($\gamma \approx 0.5772$).

If a /= 1, approximation is done using Euler-Maclaurin summation with Bernoulli numbers $B_k$: $\sum_{j=0}^{m-1} f(j) = \int_{0}^{m} f(x)dx + \sum_{k=1}^{\infty} \frac{B_k}{k!} \left[ f^{(k-1)}(x) \right]_{0}^{m}$.

So:

$$H_{m,a} = \sum_{j=1}^{m} j^{-a} = 1 + \sum_{j=2}^{m} j^{-a} = 1 + \int_{1}^{m} x^{-a} dx + \sum_{k=1}^{\infty} \frac{B_k}{k!} \left[ (x^{-a})^{(k-1)} \right]_{1}^{m} \sim 1 + \left[ \frac{x^{1-a}}{1-a} \right]_{1}^{m} = \frac{m^{1-a} - a}{1-a}$$

Note that $H_{m,0} = m$ as expected.

Following figure shows the result of the approximation for N=28 and 0<=a<=2:

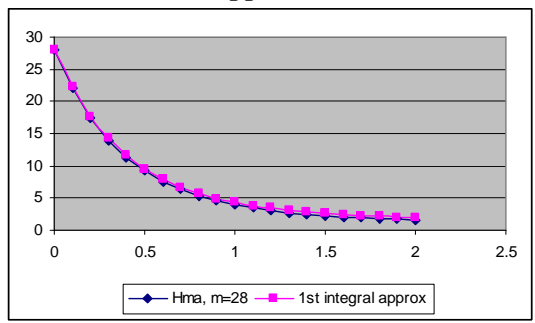

## Appendix 5: Deriving Jelenkovic a>1 Asymptotics from Fagin equations

LRU miss rate equation $MR[D] = \dfrac{N^{1-a}}{aH_{N,a}} E_{\frac{1}{a}}(E^{-1}_{1+\frac{1}{a}}(a \cdot (1 - \dfrac{D}{N})))$ means that, if D is small, D/N will tend to 0 when N increases, and consequently, Inverse function of $E_{1+1/a}$ will take its value in the vicinity of 'a' or, reciprocally, $E_{1+1/a}$ is in the vicinity of 0.
ExpInt has the following series expansion http://dlmf.nist.gov/8.19.10:

$$E_p(x) = \dfrac{1}{p-1} + x^{p-1}\Gamma(1-p) - \left(\dfrac{-x}{2-p} + \dfrac{x^2}{2!(3-p)} + ..\right)$$

Obviously when a>1, exponent p-1=1/a is less than 1 and $E_{1+\frac{1}{a}}(x)$ can be approximated by the first two terms of the series expansion in the vicinity of 0.

Hence $E_{1+\frac{1}{a}}(x) \sim a + x^{\frac{1}{a}}\Gamma(-\dfrac{1}{a})$ implying: $E^{-1}_{1+\frac{1}{a}}(x) \sim \left(\dfrac{x-a}{\Gamma(-\dfrac{1}{a})}\right)^a$.

And $E_{\frac{1}{a}}(x) = -\dfrac{d}{dx}E_{1+\frac{1}{a}}(x) = -x^{\frac{1}{a}-1}\dfrac{1}{a}\Gamma(-\dfrac{1}{a})$. Thus: $E_{\frac{1}{a}}\left(E^{-1}_{1+\frac{1}{a}}(x)\right) \sim -\left(\dfrac{x-a}{\Gamma(-\dfrac{1}{a})}\right)^{1-a}\dfrac{1}{a}\Gamma(-\dfrac{1}{a})$.

With relation $\Gamma(1+x) = x\Gamma(x)$, i.e. $\Gamma(-\dfrac{1}{a}) = -a\Gamma(1-\dfrac{1}{a})$ (http://dlmf.nist.gov/5.2.5)

$$E_{\frac{1}{a}}\left(E^{-1}_{1+\frac{1}{a}}(x)\right) \sim \left(\dfrac{x-a}{-a\Gamma(1-\dfrac{1}{a})}\right)^{1-a}\Gamma(1-\dfrac{1}{a}) = \left(\dfrac{a-x}{a}\right)^{1-a}\left(\Gamma(1-\dfrac{1}{a})\right)^a$$

Finally, with $E_{\frac{1}{a}}\left(E^{-1}_{1+\frac{1}{a}}\left(a(1-\dfrac{D}{N})\right)\right) = \left(\dfrac{D}{N}\right)^{1-a}\left(\Gamma(1-\dfrac{1}{a})\right)^a$, MR formula gives:

$$MR[D] = \dfrac{N^{1-a}}{aH_{N,a}}\left(\dfrac{D}{N}\right)^{1-a}\left(\Gamma(1-\dfrac{1}{a})\right)^a = \dfrac{1}{aH_{N,a}D^{a-1}}\left(\Gamma(1-\dfrac{1}{a})\right)^a$$

And asymptotically, when N→∞:

$$MR[D] \to \dfrac{1}{a\varsigma(a)D^{a-1}}\left(\Gamma(1-\dfrac{1}{a})\right)^a = MR_{static}[D] \cdot \dfrac{a-1}{a}\left(\Gamma(1-\dfrac{1}{a})\right)^a$$

which is the well-known Jelenkovic relation for a>1 (Jelenkovic 1999).

## *Appendix 6: Slope of WS(D) for power-law popularities*

We observe the form of the slope at the origin of the following WS (D) curves for different values of the popularity parameter: a=0, a=1 and a=2, having in mind that both axis are in logarithmic scale.

Clearly a=0 and a=1 curves have the same slope (i.e., the identity) at the origin, whereas for a=2, slope is ½.

Traces are in the range of 100M ($10^8$) and generated under the IRM hypothesis.

Note that for both a=0 and 1, the state space is 256K, where it is restricted to 21400 for a=2, even with a trace length of 1G (extending to a 256K space would have likely required a trace longer than 100G)

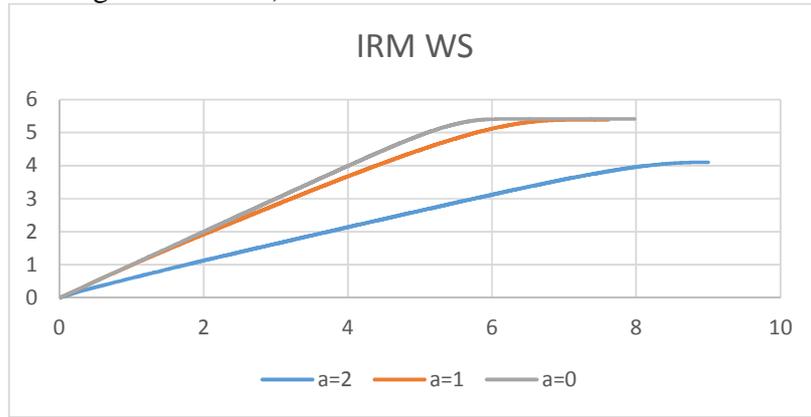

We are interested in the slope in the vicinity of 0 of WS (D) loglog representation, i.e., Y=ln WS versus X=ln D. Following computation is done using natural logarithm, however it's clearly equivalent to decimal logarithm regarding the slope at origin.

Under IRM, $WS(D) = N(1 - \frac{1}{a} \cdot E_{1+\frac{1}{a}}(\frac{D}{H_{N,a}N^a}))$, so $Y = \ln WS = \ln N + \ln(1 - \frac{1}{a} \cdot E_{1+\frac{1}{a}}(\frac{e^X}{H_{N,a}N^a}))$.

Hence $Y'(X) = \dfrac{\frac{1}{a} \cdot E_{\frac{1}{a}}(\frac{e^X}{H_{N,a}N^a}) \frac{e^X}{H_{N,a}N^a}}{1 - \frac{1}{a} \cdot E_{1+\frac{1}{a}}(\frac{e^X}{H_{N,a}N^a})}$, and, setting $Z = \dfrac{e^X}{H_{N,a}N^a}$, $Y'(Z) = \dfrac{E_{\frac{1}{a}}(Z) \cdot Z}{a - E_{1+\frac{1}{a}}(Z)}$.

When D→0⁺, X=lnD →-∞ and Z→0⁺ for N fixed.
We show that when Z→0⁺ : Y'(Z) → 1 when 0≤a≤1,
 1/a when a>1.

In order to do so, we use the series expansion of Expint in the vicinity of 0.

For p non-integer $E_p(x) = \dfrac{1}{p-1} + x^{p-1}\Gamma(1-p) - (\dfrac{-x}{2-p} + \dfrac{x^2}{2!(3-p)} + ..)$

Hence, for p=1/a non-integer, $E_{\frac{1}{a}}(x) = \dfrac{a}{1-a} + x^{\frac{1}{a}-1}\Gamma(1-\frac{1}{a}) - o(x)$ and with:

$E_{1+\frac{1}{a}}(x) = a + x^{\frac{1}{a}}\Gamma(-\frac{1}{a}) - (\dfrac{-x}{1-\frac{1}{a}}) + o(x^2)$, we finally obtain:

$$Y'(Z) = \frac{\frac{a}{1-a} + Z^{\frac{1}{a}-1}\Gamma(1-\frac{1}{a}) - o(Z)}{\frac{a}{1-a} - Z^{\frac{1}{a}-1}\Gamma(-\frac{1}{a}) + o(Z)}.$$

When Z→0⁺, this expression tends to 1 for 1/a−1>0, i.e. for a<1,

and for a>1 its limit is $\dfrac{\Gamma(1-\frac{1}{a})}{-\Gamma(-\frac{1}{a})} = \dfrac{1}{a}$.

In case 1/a=n is integer, previous series is not defined and we use the other series expansion (see previous Appendix):

$$E_n(z) = \frac{(-z)^{n-1}}{(n-1)!}(\psi(n) - \ln z) - \sum_{\substack{k=0\\k\neq n-1}}^{\infty} \frac{(-z)^k}{k!(1-n+k)},$$

Which gives (for n=1,2,3..)

$$Y'(Z) = \frac{E_n(Z) \cdot Z}{\frac{1}{n} - E_{1+n}(Z)} = \frac{\frac{-(-Z)^n}{(n-1)!}(\psi(n) - \ln Z) + \sum_{k=0,k\neq n-1}^{\infty}\frac{(-Z)^{k+1}}{k!(1-n+k)}}{\frac{1}{n} - \frac{(-Z)^n}{n!}(\psi(n+1) - \ln Z) + \sum_{k=0,k\neq n}^{\infty}\frac{(-Z)^k}{k!(n+k)}}$$

$$= \frac{\frac{-(-Z)^n}{(n-1)!}(\psi(n) - \ln Z) + \sum_{k=0,k\neq n-1}^{\infty}\frac{(-Z)^{k+1}}{k!(1-n+k)}}{-\frac{(-Z)^n}{n!}(\psi(n+1) - \ln Z) + \sum_{k=1,k\neq n}^{\infty}\frac{(-Z)^k}{k!(-n+k)}}$$

Limits when Z→0⁺ are the following:

For n =1, $Y'(Z) = \dfrac{Z(\psi(1) - \ln Z) + o(z^2)}{Z(\psi(2) - \ln Z) + o(z^2)} \to 1$

For n>=2:

$$Y'(Z) = \frac{\frac{-(-Z)^n}{(n-1)!}(\psi(n)-\ln Z) + \sum_{k=0,k\neq n-1}^{\infty}\frac{(-Z)^{k+1}}{k!(1-n+k)}}{-\frac{(-Z)^n}{n!}(\psi(n+1)-\ln Z) + \sum_{k=1,k\neq n}^{\infty}\frac{(-Z)^k}{k!(-n+k)}} = \frac{\frac{-Z}{(1-n)} + o(Z^2)}{\frac{-Z}{(-n+1)} + o(Z^2)} \to 1$$

This concludes the proof that, if a≤1, the limit of the slope at origin of WS in a loglog representation is always 1. Hence the result for any value of a>0: slope is 1 when 0<a≤1, and 1/a when a>1.

Note that, for a=0, limit of the slope at origin is also 1 since $WS(D) = N(1 - e^{-\frac{D}{N}})$, so

$Y = \ln WS = \ln N + \ln(1 - e^{-\frac{D}{N}})$. With X=lnD, $Y'(X) = Y'(D) \cdot D'(X) = \dfrac{D}{N}\dfrac{e^{-\frac{D}{N}}}{1 - e^{-\frac{D}{N}}}$,

and, setting $Z = \dfrac{D}{N}$, then: $Y'(Z) = \dfrac{e^{-Z} \cdot Z}{1 - e^{-Z}} \to 1$ when Z→0.

## *Appendix 7: Property P5: limits of F<sub>a</sub>(y) LRU/Static ratio at origin*

We use the series expansion for 1/a non-integer, $E_{\frac{1}{a}}(y) = \frac{a}{1-a} + y^{\frac{1}{a}-1}\Gamma(1-\frac{1}{a}) - o(y)$ and

$$E_{1+\frac{1}{a}}(y) = a + y^{\frac{1}{a}}\Gamma(-\frac{1}{a}) - (\frac{-y}{1-\frac{1}{a}}) + o(y^2)$$

Hence

$$F_a(y) = \frac{1-a}{a} \frac{E_{\frac{1}{a}}(y)}{1-\left(1-\frac{1}{a}E_{1+\frac{1}{a}}(y)\right)^{1-a}} = \frac{1-a}{a} \frac{\frac{a}{1-a} + y^{\frac{1}{a}-1}\Gamma(1-\frac{1}{a}) - o(y)}{1-\left(-\frac{1}{a}y^{\frac{1}{a}}\Gamma(-\frac{1}{a}) + \frac{1}{a}(\frac{-y}{1-\frac{1}{a}}) - o(y^2)\right)^{1-a}}$$

When y→0, ratio limit is the ratio of the smallest order coefficients. Thus, if 1/a-1>0 (i.e. a<1) $\lim_{y\to 0+} F_a(y) = 1$.

On the opposite, when 1/a-1<0 (a>1),

$$\lim_{y\to 0+} F_a(y) = \frac{1-a}{a} \frac{\Gamma(1-\frac{1}{a})}{-\left(-\frac{1}{a}\Gamma(-\frac{1}{a})\right)^{1-a}} = \frac{a-1}{a}\left(\Gamma(1-\frac{1}{a})\right)^a.$$

We found similar result $\lim_{y\to 0+} F_a(y) = 1$ when 1/a is integer (hence a<1) using similar reasoning as in the previous Appendix (Appendix 6).

## *Appendix 8: Property P6:* $\forall a>0, \forall y>0, F_a(y)>1$

This is equivalent to showing that:

For p>1 (0<a<1): $(1-(p-1)E_p(y))^p < (1-pE_{1+p}(y))^{p-1}$.

The direction of the inequality is reversed, that is, >, when 0<p<1 (a>1).

Relation can be checked by analyzing function $f_p(y) = \dfrac{(1-(p-1)E_p(y))^p}{(1-pE_{p+1}(y))^{p-1}}$.

One can prove that on one hand $\lim_{y\to\infty} f_p(y) = 1$ whatever p (this is direct from $E_p$ limit), and on the other hand, $\lim_{y\to 0} f_p(y) = 0$ if p>1 and $\lim_{y\to 0} f_p(y) = (1-p)^p \Gamma(1-p) > 1$ if 0<p<1 (proof is similar to P5 proof given in Appendix)

We need to verify that derivative of $f_p$ is always positive for p>1 (i.e., $f_p$ is increasing from 0 to 1) or always negative for 0<p<1 (i.e., $f_p$ is decreasing from $f_p(0)$ to 1). Noting that $\dfrac{df_p(y)}{dy}$ has the sign of $(p-1)\big(E_{p-1}(y)(1-pE_{p+1}(y)) - E_p(y)(1-(p-1)E_p(y))\big)$, for property P6 to hold, the second factor must be positive.

By using the equality $pE_{p+1}(y) = y(E_0(y) - E_p(y))$ (http://dlmf.nist.gov/8.19.12 for y>0 and p positive integer, but which can be readily extended to real positive (Chicolli, 1990)) twice, second factor is equal to the product: $\big(pE_{p+1}(y) - (p-1)E_p(y)\big)\left(\dfrac{1}{y} - E_0(y)\right)$.

First factor of product is positive, from left-hand side of $\dfrac{p-1}{p} E_p(y) < E_{p+1}(y) < E_p(y)$ inequality http://dlmf.nist.gov/8.19.E19 for y>0 and p real positive (Chicolli, 1990)). Second factor is obviously positive as well.

Consequently derivative of $f_p$ has the sign of (p-1) which completes the proof.

An interesting by-product of this inequality is:

For p=2, one gets: $(1-E_2(y))^2 < (1-2E_3(y))$ or: $(E_2(y))^2 < 2(E_2(y) - E_3(y))$.

Using $2E_3(y) = e^{-y} - yE_2(y)$, it results that: $E_2(y)(2+y-E_2(y)) > e^{-y}$.

This inequality is stronger than the well-known $E_2(y)(2+y) > e^{-y}$.

Another interesting property is: $\forall p \geq 0, \forall y \geq 0, \dfrac{E_{p-1}(y)}{E_p(y)} > \dfrac{1-(p-1)E_p(y)}{1-pE_{p+1}(y)} > 1$.

Both sides directly result from previous relations.

## Appendix 9: Property P7: $F_\infty(y) > F_1(y), \forall y > 0$

We first show that $E_0(y) > \dfrac{E_1(y)}{1-E_2(y)} > e^{E_1(y)} - 1, \forall y > 0$. From $\ln(E_0(y)+1) - E_1(y) > 0$

it stands: $E_0(y) > e^{E_1(y)} - 1$. Also from case a=1, one has: $E_0(y) > \dfrac{E_1(y)}{1-E_2(y)}$. We consider the function $f(y) = \ln\left(\dfrac{E_1(y)}{1-E_2(y)} + 1\right) - E_1(y)$. Right-hand side inequality of property P7 holds iff f(y)>0, $\forall y>0$.

One can see that f(y) tends to 0 when y→∞, a positive number when y→0 (which can be evaluated to $\gamma - \ln(1-1/\gamma)$, using $E_1$ and $E_2$ series expansion at y=0 and series expansion of $\ln(1+X) = -\ln(1/X) + 1/X + o(1/X^2)$ in the vicinity of X=+∞), and its derivative has the sign of $E_0(y)(1-E_2(y))(E_1(y) - E_2(y)) - (E_1(y))^2$. In turn this expression can be proven always negative for y real positive, since it tends to -1 when y→0, tends to 0 when y→∞, and its derivative can be brought down to $E_0(y)(xE_2(y)(1-E_2(y)) + E_1(y)(x + e^{-x} - 1))$ which is always positive (each term is positive).

**Property P7**: $F_\infty = \dfrac{E_0(y)}{e^{E_1(y)} - 1} > \dfrac{-E_1(y)}{\ln(1-E_2(y))} = F_1, \forall y > 0$ .

This stems from a=1 analysis. When y increases from 0, $F_1$ increases from 1 to a maximum where y is solution $y_0$ of $F_1(y) = \dfrac{(1-E_2(y))E_0(y)}{E_1(y)}$. From previous property, it holds that $F_\infty(y) > \dfrac{E_0(y)(1-E_2(y))}{E_1(y)}$ hence $F_\infty(y_0) > F_1(y_0)$. Obviously $F_\infty(y) > F_1(y)$ for

y<$y_0$ since $F_\infty$ is decreasing and $F_1$ increasing. For y>$y_0$, $\dfrac{E_0(y)(1-E_2(y))}{E_1(y)} > F_1(y)$